%% file: proof-of-contact.tex
\documentclass[letterpaper,twocolumn,10pt]{article}
\usepackage{graphics}
\usepackage{graphicx}
\usepackage{amsmath}
\usepackage{amsthm}
\usepackage{amsfonts}
\usepackage{amssymb}
\usepackage{color}
\usepackage{subfigure}
\usepackage{xspace}
\usepackage{algorithm}
\usepackage{algcompatible}
\usepackage{algpseudocode}
\usepackage{breqn}
\usepackage{cite}
\usepackage{multirow}

\usepackage{fancyhdr}
\usepackage{kantlipsum}
\usepackage{eso-pic}
\usepackage[%
  pdfborder={0 0 0},%
  bookmarksdepth=subsection,%
  bookmarksnumbered,%
  bookmarksopen,%
]{hyperref}
\algrenewcommand\algorithmiccomment[1]{// {\itshape #1}}

\newcommand{\FBB}{\mathbb{F}}

\newcommand{\PK}{\mathsf{pk}}
\newcommand{\VK}{\mathsf{vk}}
\newcommand{\RC}{\mathcal{R}_C}

\newcommand{\LC}{\mathcal{L}_C}
\newcommand{\ZIN}{\ensuremath{\vec{z_{in}}}}

\newcommand{\ZOU}{\ensuremath{z_{out}}}
\newcommand{\PIN}{\ensuremath{\vec{\pi_{in}}}}
\newcommand{\POU}{\ensuremath{\pi_{out}}}
\newcommand{\ZLO}{\ensuremath{z_{\mathsf{loc}}}}

\newcommand{\pred}{\ensuremath{\prod}}

\title{SNARKs to the rescue: proof-of-contact in zero knowledge}
\author{
Zachary Ratliff and Joud Khoury\\
Raytheon BBN Technologies 
}
\date{} 

\begin{document}

\maketitle

\abstract{
This paper describes techniques to help with COVID-19 automated contact tracing, and with the restoration efforts.
We describe a decentralized protocol for ``proof-of-contact'' in zero knowledge where a person can publish a short cryptographic proof attesting to the fact that they have been infected and that they have come in contact with a set of people without revealing any information about any of the people involved. 
More importantly, we describe how to compose these proofs to support broader functionality such as proofs of $n$th-order exposure which can further speed up automated contact tracing.
The cryptographic proofs are publicly verifiable, and places the burden on the person proving contact and not on third parties or healthcare providers rendering the system more decentralized, and accordingly more scalable. 
}

\input{intro}

\input{background}

\input{protocol}


\input{pcd}

\input{implementation}

\input{performance}

\input{related}

\bibliographystyle{IEEEtran}
\bibliography{refs}

\end{document}

%% file: intro.tex
\section{Introduction}

Contact tracing, identifying and notifying individuals who have been in close contact with an infected person, is widely recognized as an essential tool in protecting against the spread of the novel COVID-19 virus.
Automated approaches to contact tracing can help significantly scale the effort relative to manual approaches alone which tend to be slower and more labor intensive.
Implementations of automated contact tracing systems must however address the privacy concerns of individuals in order to enjoy widespread adoption, something that early straightforward attempts failed to do~\cite{singapore2020, cho2020contact}.

There is a large body of recent proposals for automated Bluetooth-based contact tracing systems with differing privacy guarantees.
We refer the reader to~\cite{Vaudenay:2020} for a recent survey.
The systems fall into two general categories based on their information flows: decentralized vs centralized.
With decentralized approaches, a user's mobile device generates ephemeral randomized tokens that are regularly broadcast to (and received by) nearby mobile devices.
Devices save the tokens they broadcast and the ones they receive for a defined period of time.
Once an individual tests positive, she can opt to report all the tokens her application generated.
Reporting is done with the help of the healthcare provider or some third party.
Other individuals who saw the token, and accordingly were in close physical proximity to the infected individual, learn they are at-risk and may seek testing and/or quarantine as a result. 
Centralized approaches are similar, except a central server generates the ephemeral tokens that users share with each other and reporting involves the central server making the connections and alerting users. 

The majority of these existing proposals for automated contact tracing are slow to react and do not adequately address {\em exposure risk}.
Specifically, they only alert the first level of individuals who have come in close contact with the infected individual after the latter tests positive.
Such first order contact tracing may not be fast enough to control the spread of the virus in a timely manner given that there is a period of time in which individuals can be asymptomatic but contagious, and this period could be longer than the virus incubation period.
Consider for example the following scenario.
Alice is asymptomatic but contagious at time $t_0$ and comes in contact with Bob.
Bob gets infected and comes in contact with Charlie at time $t_1\geq t_0 + P_I$ where $P_I$ is the virus incubation period.
Alice starts showing symptoms and tests positive at time $t_2\geq t_1$, at which point Bob gets notified.
Bob may not show symptoms, may wait to get tested, or may not even get tested.
Even if Bob gets tested at time $t_3 > t_2$, there is a period of time (could be several days) during which Charlie is not even aware of the exposure risk, and is going about his business as usual.

We propose a new protocol for privacy-preserving contact tracing that does not suffer from these limitations. 
Our protocol permits an individual A who tests positive to quickly furnish a cryptographic proof attesting to the the following statements:
\begin{enumerate}
\item individual A was in close proximity to individual B at some time $t$
\item individual A tested positive for the virus at time $t'$
\item $t'$ is within $x$ days of $t$
 \end{enumerate}
in zero knowledge i.e., without leaking information about A or B.
A produces and publishes the proof. 
Anyone, including, B can publicly verify the proof, and seek testing if the proof checks and they are involved.
Using this first proof, individual(s) B who came in close proximity to A can then quickly publish a cryptographic proof attesting to the fact that B was in close proximity to the individual who tested positive (in this case A), and that B was in close proximity to other individual(s) C at $t'' \geq t + P_I$. 
This allows individual(s) C who came in close proximity to B to realize their exposure risk in a timely manner, and act accordingly.

Our protocol relies on zero-knowledge succinct non-interactive arguments of knowledge (zkSNARKs) as the cryptographic building block. 
The protocol builds on existing Bluetooth-based decentralized approaches and allows clients to provide a cryptographic proof of proximity after a positive diagnosis. 
These cryptographic proofs are \emph{succinct}, consisting of only a few hundred bytes, and take just a few milliseconds to verify. 
The zero knowledge property ensures that a person verifying these proofs only learns the statement ``I was close to someone who tested positive for the virus'' or ``I was close to someone who was close to someone who tested positive for the virus'' and so on, but nothing else such as who the person is or where the interaction occurred (with some caveats discussed later in the paper). Our approach is fully decentralized requiring little assistance from the healthcare provider, and may be extended to support broader functionality. 
We start by describing the simple proof-of-contact protocol, and extend it to support $n$th-order exposure by composing proofs of contact using proof-carrying data (PCD)~\cite{chiesa2012proof}. 

In summary, our protocol offers the following benefits:
\begin{itemize}
\item Efficiency is achieved using an efficient pre-processing zkSNARK construction and performing the signature verification outside the SNARK to reduce prover cost~\cite{naveh2016photoproof}. In the case of $n$th-order exposure proofs, signature verification occurs inside the zkSNARK, however, an arithmetic circuit friendly digital signature scheme is used to reduce circuit size. 
\item No trusted third parties or databases required. The public registry need not be trusted. We only require that the zkSNARK for the desired functionality is correctly setup.
\item Strong end-to-end privacy guarantees. Proximity tokens are not shared; not with third parties nor with healthcare providers.
\item Correctness. A valid proof guarantees the authenticity of the user's test results and the validity of the statement.
\item Adoption/Practicality. A medical organization only needs to sign records using an existentially unforgeable and publicly verifiable signature scheme. This is a simple task for the medical organization and it deters malicious (non-infected) users from seeking signatures.
\item Decentralization. The burden is on the infected person to actually prove and publish. This allows better scaling (instead of requiring providers or third parties to centrally manage patient proximity data) and potentially better privacy since the user (the stakeholder) has full control over their private data and can share at will.
\item $n$th-order exposure notifications. $n$th-order exposure risk is available in a timely manner through proof composition. This allows users who may have had secondhand contact (or beyond) with an infected individual, to learn in zero-knowledge about this exposure. 
\end{itemize}

%% file: background.tex
\section{Background}
\subsection{Proximity Tokens}
Contact tracing requires monitoring and recording physical interactions between clients. For example, if Alice walks into a cafe where Bob is eating, a method for detecting and measuring their proximity is needed. There have been several works proposing various means of proximity sensing between mobile phones, including using Bluetooth \cite{liu2013face}, WiFi \cite{sapiezynski2017inferring}, and audio \cite{thiel2012sound} signals.

Regardless of the underlying technology, we assume a mobile phone frequently broadcasts {\em proximity tokens} that are received by nearby phones. For example, within each {\em epoch}, the phone frequently broadcasts its unique token, and receives tokens from nearby phones. This simplified model has been adopted by the majority of decentralized privacy-preserving contact tracing protocols~\cite{Vaudenay:2020}.

\subsection{Preprocessing zkSNARK}
We review the definitions of arithmetic circuits, preprocessing zero knowledge succinct non-interactive arguments of knowledge (pp-zk-SNARKs) and we refer the reader to~\cite{ben2017scalable} for details.

First, we introduce arithmetic circuit satisfiability in Field $\FBB$. 
An $\FBB$-arithmetic circuit $C:\FBB^n \times \FBB^h \rightarrow \FBB ^l$ is defined by the relation $\RC=\{(x, a): C(x, a)=0^l\}$. Here $a$ is called the witness (auxiliary input) and $x$ is the public input and the output is $0^l$.  The language of the circuit is defined by $\LC=\{x:\exists a, C(x,a)=0^l\}$.
Here $x \in \FBB^n$ (i.e., $x$ is represented as $n$ field elements), $a \in \FBB^h$, and the output in $\FBB^l$. 

A hashing circuit for example takes the (private) input/witness $a$ and its hash $x$, and asserts that $H(a)=x$.

A preprocessing zkSNARK for $\FBB$-arithmetic circuit satisfiability comprises three algorithms $(G, P, V)$, corresponding to the {\em Generator}, the {\em Prover}, and the {\em Verifier}. 
\begin{description}
\item[$G(\lambda, C) \rightarrow (\PK, \VK)$] Given a security parameter $\lambda$ and the $\FBB$-arithmetic circuit $C$, sample a keypair comprising a public proving key $\PK$ and a public verification key $\VK$.
\item[$P(\PK, x, a) \rightarrow (\pi)$] Given the public prover key $\PK$ and any $(c,a) \in \RC$, generate a succinct proof $\pi$ attesting that $x \in \LC$
\item[$V(\VK, x, \pi) \rightarrow b \in \{0,1\}$] checks that $\pi$ is a valid proof for $x \in \LC$.
\end{description}

\subsection{Proof-carrying data}
Proof-carrying data (PCD) captures the security guarantees necessary for recursively composing zkSNARKs. More specifically, given a compliance predicate $\pred$, a PCD system checks that a local computation involving a set of incoming messages $\ZIN$, private local data $\ZLO$, and outgoing message $\ZOU$, is $\pred$-compliant.

Formally, a proof-carrying data system consists of three polynomial-time algorithms $(G, P, V)$ corresponding to the \emph{Generator}, \emph{Prover}, and \emph{Verifier}. 

\begin{description}
\item $G(\lambda, \pred) \rightarrow (\PK, \VK)$ Given a security parameter $\lambda$ and the compliance predicate $\pred$ expressed as a $\FBB$-arithmetic circuit, sample a keypair comprising a public proving key $\PK$ and a public verification key $\VK$.
\item $P(\PK, \ZIN, \PIN, \ZLO, \ZOU) \rightarrow (\ZOU, \POU)$ Given the public prover key $\PK$, a set of input messages $\ZIN$ along with compliance proofs $\PIN$, local input $\ZLO$, and output $\ZOU$, generate a succinct proof $\POU$ attesting that $\ZOU$ is $\pred$-compliant.
\item $V(\VK, z, \pi) \rightarrow b \in \{0,1\}$ checks that $\ZOU$ is $\pred$-compliant.
\end{description}

%% file: protocol.tex
\section{Proof-of-Contact Protocol}\label{sec:protocol}
Consider an existentially unforgeable signature scheme $\mathcal{S}=(G_S, S_S, V_S)$ (e.g., ECDSA) with private signing key $v_s$ and public verification key $p_s$.
Let $H, H_1, H_2$ be three collision-resistant hash functions, and
let $(G, P, V)$ be a pp-zk-SNARK.
The baseline protocol builds on~\cite{canetti2020anonymous} and works as follows:
\begin{itemize}
\item Trusted setup phase: a trusted entity sets up the system and runs the generator algorithm $G(\lambda, C) \rightarrow (\PK, \VK)$; we describe the circuit $C$ in more detail shortly. During this phase, each healthcare provider obtains a certificate for its signing key signed by a trusted certification authority.
\item Each user generates a private random string $S$
\item User A generates a random token every time period $t$ (the {\em epoch} e.g., 5 minute intervals) as $T_{A,t}=H_1(S, t)$, and frequently broadcasts the token. We omit the time subscript hereafter whenever it is clear.
\item Whenever user A receives a proximity token from user B at time $t$, she computes $h=H_2(T_A, T_B, t)$ and stores it for 14 days. User B computes the same output. Here we sort the tokens (e.g., lexicographically) before passing them to the hash function.
\item  User A tests positive for the virus at time $t'$, and obtains a ``COVID.positive'' test result from a medical provider.  User A computes $h_s=H(S, \mathsf{COVID.positive}, t')$ and requests signature $s=S_S(v_s, h_s)$ from the healthcare provider where $v_s$ is the provider's private signing key.
Note that user A does not have to reveal her secret $S$ to the provider. 
User A may provide $h_s$ only, and a cryptographic proof that $h_s=H(S, \mathsf{COVID.positive}, t')$ for some valid private witness $S$.
\item User A then generates a short cryptographic proof using $P(\PK, (h, h_s), (S, T_A, T_B, t')) \rightarrow \pi$ attesting to these facts
\begin{enumerate}
\item  $h_s=H(S, \mathsf{COVID.positive}, t')$
\item $T_A=H_1(S, t)$
\item $h=H_2(T_A, T_B, t)$
\item $t'-t \leq 14$ days
\end{enumerate}
\item User A publishes tuple $(\pi, h, h_s, s)$ to some public registry. If the public registry already contains a tuple with the value $h$, then the user does not upload these values (in order to prevent linkability). Several techniques may be used here for network unlinkability (e.g., the user app can either use mixing or onion routing solutions, or the provider can publish the material on behalf of the user). 
\item User B checks the public registry periodically to find a matching $h$ and can quickly verify the proof using $V(\VK, (h, h_s), \pi)$. If the proof checks, user B verifies the signature $V_S(p_s, s, h_s)$ given $h_s$ and the public verification key $p_s$ of the healthcare provider. 
\item User B seeks testing, and can show the proof-of-contact to her healthcare provider to expedite the process if needed.
\end{itemize}

\subsection{Security Analysis}
\begin{description}
\item [Linkability] Tokens are never shared, or published. Only the hash of two tokens is published after a user tests positive. This means different tokens may not be linked as belonging to the same user. The same is true with linking different hashes. Recall when reporting a positive test, user A publishes $h=H_2(T_A, T_B, t)$ for all proximity edges. Only user B or some dishonest user C who forms a clique with A and B at time $t$ may learn $h$. Since user C is part of the clique, $h$ does not leak additional information. User C cannot use $h$ to create valid proofs on behalf of A or B without knowledge of their private strings $S$.
\item[Identification] After seeing a proof containing $h$, a curious user B who keeps track of all physical encounters can {\em a posteriori} identify the infected person in some form. This attack is common to the majority of the decentralized systems~\cite{Vaudenay:2020}. We observe that some form of this leakage is inherent to the protocol. For example, if user B has only encountered one person before getting alerted, user B will be able to identify the infected person no matter how privacy-preserving the alert/protocol is. This may be acceptable in some cases, for example, learning that the ``tall person in the dairy aisle at the grocery store'' tested positive.
A more recent decentralized protocol that mitigates identification attacks has been proposed using ``parroting''~\cite{cryptoeprint:2020:863}. We believe the general idea presented in this paper can be applied to this class of protocols as well.
\end{description}




%% file: pcd.tex
\section{Transitive exposure proofs}
As discussed earlier, it can be beneficial to provide more granular $n$th order exposure risk data to users to limit the spread of the virus. 
For example, a user may want to know whether they have had \emph{transitive exposure} to a virus. 
Consider that Alice comes in contact with both Bob and Charlie independently of one another. Later, Bob tests positive for the virus, and Alice is alerted that she is at risk. Although Charlie did not directly come in contact with a carrier of the virus, he may find it useful to know that someone he came in contact with has. This transitive approach to contact tracing could enable more informative statistics for users such as a risk profile, i.e., a risk score based on how many degrees of exposure an individual has. Someone who is four transitive hops away from a virus carrier would be at lower risk from someone who is two hops away. 

A strawman approach to extending the proof-of-contact protocol for transitive proofs works as follows:
\begin{itemize}
    \item As in the original protocol, a trusted entity sets up the system and runs the generator algorithm $G(\lambda, C_2) \rightarrow (\PK_2, \VK_2)$; here $C_2$ is an additional circuit with corresponding prover and verifier keys $(\PK_2, \VK_2)$, for proving transitive exposure. 
    \item User B checks the public registry periodically to find a matching $h_i$ (from some user A who tested positive) and can quickly verify the proof using $V(\VK, (h_i, h_s), \pi)$. If the proof checks, user B verifies the signature $V_S(p_s, s, h_s)$ given $h_s$ and the public verification key $p_s$ of the healthcare provider. 
    \item User B then generates a short cryptographic proof using $P(\PK_2, (h_i, h_j), (S, T_A, T_{B1}, T_{B2}, T_C)) \rightarrow \pi$ attesting to these facts
\begin{enumerate}
\item $h_i=H_2(T_A, T_{B1}, t_1)$
\item $h_j=H_2(T_{B2}, T_C, t_2)$
\item $t_2 - t_1 \leq 3$ days
\end{enumerate} 
\item User B publishes tuple $(\pi, h_i, h_j)$ to the public registry. 
\item User C checks the public registry periodically to find a matching $h_j$ and can quickly verify the proof using $V(\VK_2, (h_i, h_j), \pi)$. If the proof checks, user C can recursively verify the next proof in the chain until eventually arriving at the original proof. Finally, user C verifies the original proof using $V(\VK, (h_i, h_s), \pi)$.
\end{itemize}

Observe that in this case the zkSNARK includes the constraint $t_2 - t_1 \le 3$, corresponding to the $3$ day incubation period of COVID-19. This parameter is configurable, however, in general the time that Bob comes in contact with Charlie should come after the time Bob came in contact with Alice plus the incubation period. This will reduce the number of false positives that arise when Bob alerts Charlie of 2nd-order exposure even though Bob could not have possibly become contagious from Alice yet. 

\subsection{Transitive exposure using proof-carrying data}
The above protocol suffers from a linkability flaw with the uploaded $(h_i, h_j)$ pairs. An adversary observing the public registry can deduce that whoever uploaded the tuple $(\pi_1, h_i, h_s, s)$ must have came in contact with the person who uploaded the tuple $(\pi_2, h_i, h_j)$, since $h_i$ is present in both tuples. In order to circumvent this drawback, we modify the protocol to use proof-carrying data (PCD). Using PCD, previous proofs in the chain are verified and a proof that this verification was performed correctly is provided. The PCD system hides the details of intermediate proofs, while allowing a user to verify that the entire chain is valid. Instead of uploading the pairs $(h_i, h_j)$, transitive proofs consist only of single $h$ values which are indistinguishable from random.

For proof-of-contact, we represent the compliance predicate $\pred$ as the hospital signature verification  algorithm $V_S(p_s, s, h_s)$, coupled with the steps necessary to prove that the randomness of $h_i$ is consistent with the randomness of some $h_j$. More formally, a user who tested positive can perform the $\pred$-compliant computation $M_0$ that takes as input $\ZIN = (h_s, s, p_s)$, $\ZLO = (S, t,t',T_A, T_{B1})$ and outputs $h_i$ satisfying the following constraints:

\begin{enumerate}
    \item $h_s = H(S_A, \mathsf{COVID.positive}, t')$
    \item $T_A=H_1(S, t)$
    \item $h_i = H_2(T_A, T_{B1}, t)$
    \item $t'-t \leq 14$ days
    \item $V_S(p_s, s, h_s) = 1$
\end{enumerate}

The user then uploads the value $h_i$ along with a cryptographic proof attesting that $M_0$ is $\pred$-compliant. 

For proving transitive exposure, a user B who sees the value $h_i$ along with the PCD proof $\pi_i$ attesting to first-hand exposure, performs the $\pred$-compliant computation $M_1$ that takes as input $\ZIN = (h_i, \pi_i)$, $\ZLO = (t_1, t_2, T_A, T_{B1}, T_{B2}, T_C)$ and outputs $h_j$ satisfying the following constraints:
\begin{enumerate}
    \item $h_i = H_2(T_A, T_{B1}, t_1)$
    \item $h_j = H_2(T_{B2}, T_C, t_2)$
    \item $t_2- t_1 \leq 3$ days
\end{enumerate}

\begin{figure}
    \centering
    \includegraphics{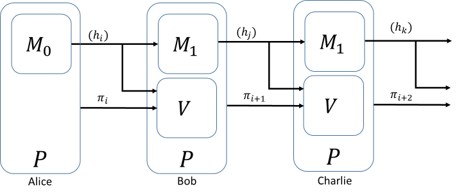}
    \caption{Overview of proof-carrying data for transitive exposure}
    \label{fig:pcd-diagram}
\end{figure}

Additionally, user B runs a verifier circuit over $\pi_i$ and provides a cryptographic proof that $V(\VK, h_j, \pi_i) = 1$ and $M_1$ is $\pred$-compliant. Figure \ref{fig:pcd-diagram} illustrates the complete flow from proof-of-contact to proof of transitive exposure.

\subsection{Proofs of surface transmission via PCD}
In some cases, contact tracing by measuring proximity between users may not be sufficient for effectively curbing the spread of a virus. A virus that lives for extended periods on surfaces could transmit from one user to another even though they have never been in close contact. For example, if a contagious user Alice sits on a park bench, Bob, who visits the park the next day, may become infected from sitting on the same bench. If Alice tests positive, it would be ideal that users who are at risk from the surface spread of the virus are alerted.

One approach is to place Bluetooth devices around public spaces, and have them participate in the contact tracing protocol. The devices could exchange tokens with users and verify proofs in the usual way. After discovering a matching token in the public registry, and verifying the corresponding proof, the device uploads a transitive proof of exposure, which alerts users of the surface transmission risk. 

Suppose rather than using PCD, the Bluetooth device on the park bench simply uploads its secondary tokens after a user Alice tests positive, i.e., the tokens exchanged with other users within 14 days of Alice's park visit. Although these users are alerted of surface contact risk, they must trust that the park bench device is acting honestly since there is no way of verifying that Alice actually came in contact with the park bench. By using PCD, the transitive proofs maintain the security and privacy guarantees from the single-hop contact tracing protocol.

%% file: implementation.tex
\section{Implementation challenges}
In this section we describe the various challenges associated with implementing our approach in a real-world system.
\subsection{Anonymization of network traffic}
A passive adversary that has the ability to view a large portion of network traffic may de-anonymize users as they interact with the public registry. For instance, if a user uploads their proofs from a home router, an adversary may be able to determine which individual tested positive based on network traffic analysis alone. 

Similar to other works \cite{canetti2020anonymous}, we propose using Tor for efficient anonymization of the uploading/downloading of proofs. For more stringent privacy guarantees (such as those which would thwart a nation state adversary), techniques from metadata private messaging protocols could be used \cite{angel2016unobservable}\cite{corrigan2015riposte}\cite{kwon2020xrd}\cite{kissner2004private}\cite{van2015vuvuzela}\cite{lazar2018karaoke}. 

\subsection{Trusted setup}
A big question with pre-processing zero-knowledge SNARKs, is which entity performs the trusted setup (generator) phase. One approach is for several community organizations to perform a secure multi-party computation (MPC) protocol. For example, the World Health Organization (WHO), Massachusetts General Hospital (MGH), and National Institute of Health (NIH) could jointly compute the trusted-setup, and users would have high confidence in the systems security - as long as they trust that these parties will not collude with one another. 

Alternatively, we could instantiate the system with a \emph{transparent} zero-knowledge SNARK scheme such as \cite{chiesa2020fractal}\cite{bowe2019halo}. These zkSNARKs do not require a trusted setup, however, this benefit comes at the cost of larger proof sizes.

\subsection{Choice of digital signature scheme for transitive exposure proofs} Enabling $n$th-order exposure notifications requires encoding a digital signature scheme inside the compliance predicate. Without careful consideration of the scheme used, this can significantly increase the size of the compliance predicate, resulting in a prohibitive proving time. For this reason, we choose the RSA digital signature scheme which can be represented efficiently over $\FBB_p$ by choosing public exponent $e = 3$ and performing modular multiplication via radix $\lfloor \sqrt{p} \rfloor$ arithmetic as suggested by Naveh and Tromer \cite{naveh2016photoproof}. 

%% file: performance.tex
\section{Performance Evaluation}
We implemented a simplified proof-of-contact zkSNARK (without recursive composition) using the \emph{libsnark} library \cite{libsnark}. The library uses the NP-complete language R1CS to express the arithmetic circuits representing the zkSNARK. The libsnark library provides existing R1CS \emph{gadgets} for performing useful functionality, such as comparisons and collision-resistant hashing. Additionally, it includes an implementation of the subset-sum collision-resistant hashing gadget, which we use as an efficient one-way hash.

We characterize the performance of our proof-of-contact zkSNARK in terms of the running time and key sizes for both the prover and verifier (Table~\ref{tbl:perf}). Since the generator phase is only executed once during setup, we provide concrete numbers on the size of the arithmetic circuit (3060 gates) but disregard the time of the generator (166 ms). The circuit did not account for sorting.

\begin{table}[th]
\centering
\caption{Performance of pp-zk-SNARK implementation on MacBook Pro with 2.9 GHz Intel core i9 and 32 GB RAM}
\label{tbl:perf}
\begin{tabular}{c|c|c|}
\cline{2-3}
                                            & \textbf{Prover} & \textbf{Verifier} \\ \hline
\multicolumn{1}{|c|}{\textbf{Running time} (ms)} & 65            & 9               \\ \hline
\multicolumn{1}{|c|}{\textbf{Key size} (KB)}     & 722           & 30              \\ \hline
\end{tabular}
\end{table}

%% file: related.tex
\section{Related Work}
There has been a flurry of work on privacy-preserving automated contact tracing  \cite{altuwaiyan2018epic}\cite{rivest2020pact}\cite{canetti2020anonymous}\cite{cryptoeprint:2020:863}\cite{troncoso2020decentralized}\cite{pepppt2020}\cite{applegoogle2020}\cite{berke2003assessing}\cite{raskar2020apps}\cite{chan2020pact}\cite{trieu2020epione}. Although most of these works suggest similar techniques for estimating and exchanging proximity information between users, the underlying cryptographic protocols and their privacy guarantees differ. 

\cite{canetti2020anonymous}\cite{rivest2020pact}\cite{pepppt2020} use randomly generated pseudonyms that nearby users can exchange over Bluetooth. Individuals who test positive for a virus can upload their generated pseudonyms to a public registry, allowing other users to match the tokens they have collected with those in the registry. The authors suggest that healthcare workers should be the ones to upload users' tokens to the public registry after giving a positive test diagnosis in order to prevent malicious polluting of the database. Similar to the protocol introduced in this work, mixing can be applied to prevent linkability via traffic analysis. 

Apple and Google have released a protocol specification \cite{applegoogle2020} that closely resembles that of \cite{canetti2020anonymous}\cite{rivest2020pact}\cite{pepppt2020}. Users generate a rolling pseudorandom identifier and some associated encrypted metadata, that nearby users exchange over Bluetooth. The pseudorandom identifiers are derived using the current time and temporary exposure keys, which get distributed after a positive diagnosis. 

\cite{berke2003assessing} proposes partitioning GPS and time data into discrete spatiotemporal points and obfuscating these points using a one-way hash function. Infected users upload their obfuscated location histories after redacting personally identifiable information such as the GPS coordinates that represent a home or work address. Using private-set intersection (PSI), individuals can privately determine whether or not their location history overlaps with that of infected users. 

More recently, \cite{cryptoeprint:2020:863} performs a security analysis of existing contact tracing approaches and groups \cite{altuwaiyan2018epic}\cite{applegoogle2020}\cite{canetti2020anonymous}\cite{rivest2020pact}\cite{chan2020pact}\cite{troncoso2020decentralized} together under the "upload what you've sent" paradigm, or \emph{ReBabblers}. These approaches are overall less secure than a "upload what you've heard" (\emph{parroting}) alternative presented in \cite{cryptoeprint:2020:863}, which prevents inverse sybil attacks \cite{Vaudenay:2020}\cite{chan2020pact}. Although the scheme presented in this paper falls under the ReBabbler paradigm, we note that producing individual proofs of contact requires work on behalf of the person reporting infection. This is especially true for proofs of $n$th-order exposure, where the cost of producing a proof is further amplified. The computational burden of uploading many proofs of contact may discourage sybil attacks while adhering to the "upload what you've sent" architecture. Additionally, while we our protocol is built upon the ``rebabbling'' variant, 
we believe that the general idea is still applicable to the more private ``parroting'' variant.
We leave this for future work.

All of the approaches described above provide different flavors of privacy and decentralization. However, each solution places an increased burden on the healthcare providers relative to the zero-knowledge SNARK technique we have outlined. Our approach requires only that healthcare workers sign positive diagnoses rather than generate one-time codes or upload tokens to a public registry. Additionally, our approach is fully decentralized and supports broader functionality such as proofs of $n$th-order exposure, not currently supported by the other proposed solutions. 